\documentclass[a4paper]{spie}  %>>> use this instead for A4 paper
\usepackage[]{graphicx}

\title{In-orbit performance of the EPIC-MOS detectors on XMM-Newton} 

\author{S. Sembay\supit{a}, A. Abbey\supit{a}, B. Altieri\supit{b}, R. Ambrosi\supit{a}, D. Baskill\supit{a}, P. Ferrando\supit{c}, K. Mukerjee\supit{a}, 
\skiplinehalf
A. Read\supit{a} and M. J. L. Turner\supit{a}
\skiplinehalf
\supit{a}Department of Physics, Leicester University, UK; \\
\supit{b}XMM-SOC, ESA/Vilspa, Spain; \\
\supit{c}CEA/Sap, Saclay, France
}

\authorinfo{Further author information: (Send correspondence to S. Sembay)\\S. Sembay: E-mail: sfs5@star.le.ac.uk, Telephone: 44 116 252 3507}
\begin{document} 
  \maketitle 

%%%%%%%%%%%%%%%%%%%%%%%%%%%%%%%%%%%%%%%%%%%%%%%%%%%%%%%%%%%%% 
\begin{abstract}

XMM-Newton was launched into space on a highly eccentric 48 hour orbit on 
December 10th 1999. XMM-Newton is now in its fifth year of operation and has
been an outstanding success, observing the Cosmos with imaging, spectroscopy 
and timing capabilities in the X-ray and optical wavebands. The EPIC-MOS
CCD X-ray detectors comprise two out of three of the focal plane instruments on
XMM-Newton. In this paper we discuss key aspects of the current status and 
performance history of the charge transfer ineffiency (CTI), 
energy resolution and spectral redistribution function (rmf) of EPIC-MOS in 
its fifth year of operation.

\end{abstract}

%>>>> Include a list of keywords after the abstract 


%%%%%%%%%%%%%%%%%%%%%%%%%%%%%%%%%%%%%%%%%%%%%%%%%%%%%%%%%%%%%
\section{INTRODUCTION}
\label{sect:intro}  % \label{} allows reference to this section

XMM-Newton\cite{Jansen01} carries three telescopes which combined provide the 
highest throughput of any X-ray observatory launched to date. At the focus of 
each telescope is a CCD-based imaging spectrometer, the trio making up the 
European Photon Imaging Camera (EPIC). Two of the cameras (EPIC-MOS) use 
seven EEV type 22 MOS CCDs\cite{Turner01} each whilst the third (EPIC-pn) 
uses twelve pn CCDs\cite{Struder01}. 

The design of the EPIC detectors must account for the radiation environment 
in space in order to minimise radiation damage to the CCDs. 
The EPIC-MOS cameras have shielding designed to block the most damaging 
particles: non-relativistic protons (energies below 30~MeV) which have a high 
displacement cross-section. Electron divertors in each telescope prevent these
particles reaching the focal plane. CTI in CCDs is temperature dependent and 
EPIC was designed with reserve cooling 
power so that the CCDS could be operated at an optimum temperature of around 
$-120^{\circ}$C. Finally, the option to anneal (heat) the CCDs to repair 
damage is available.

\section{The EPIC-MOS Cameras in Orbit}

\subsection{Charge Transfer Inefficiency}
\label{charge-transfer-inefficiency}

Radiation damage increases the process of CTI within CCDs by displacing 
atoms in the silicon lattice and thereby 
creating sites of locally low potential. These sites are charge traps which 
reduce the magnitude of charge packets as they are transferred down to the 
readout node. The total charge lost from a given X-ray is approximately a 
linear function of the number of transfers from pixel to pixel. CTI also 
degrades the energy resolution of the CCDs because of statistical 
broadening. This is described in Section~\ref{sect:resolution}.

\begin{figure}
   \begin{center}
   \begin{tabular}{c}
   \includegraphics[width=6cm,height=8.3cm,angle=90]{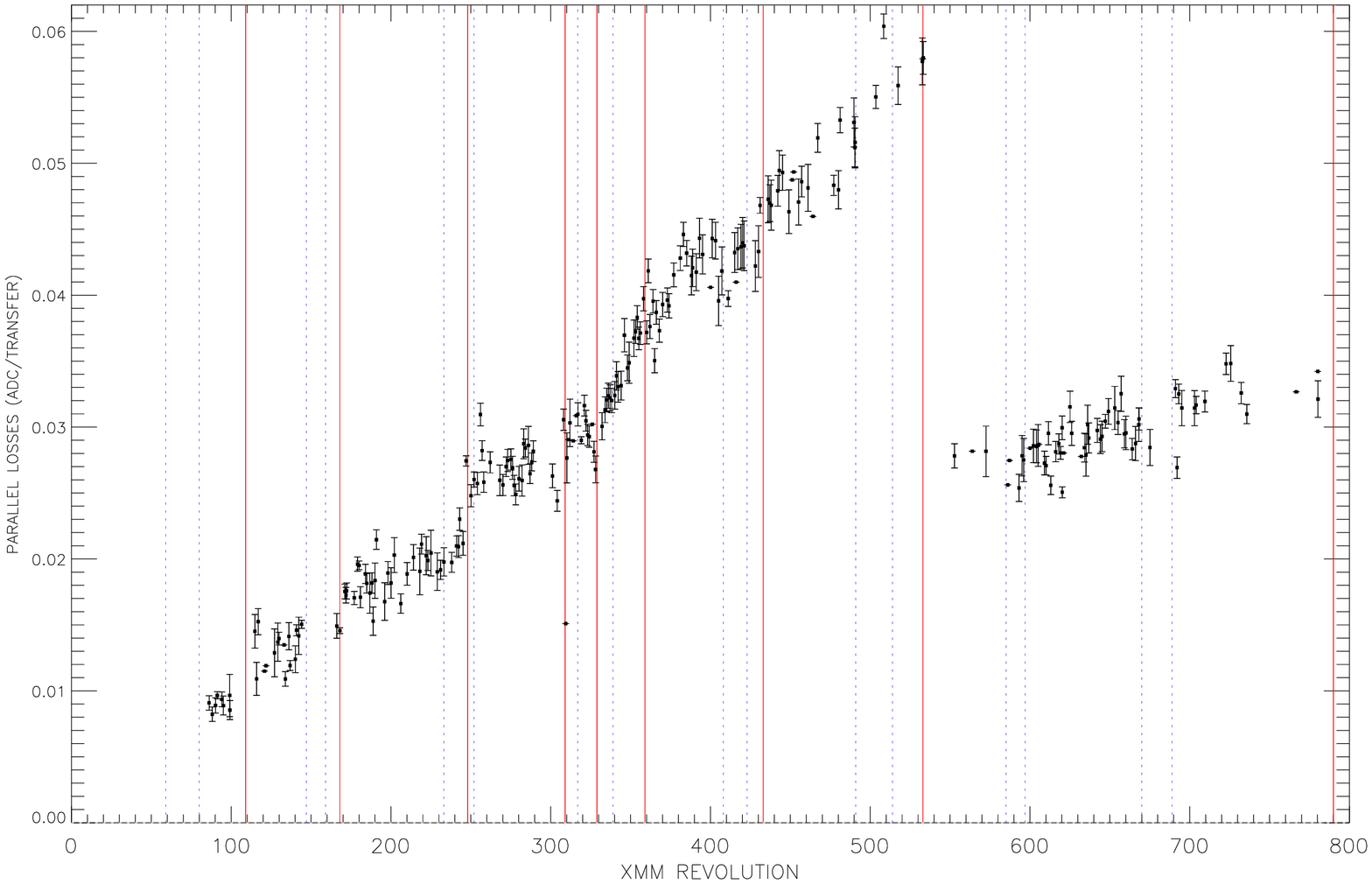}
   \includegraphics[width=6cm,height=8.3cm,angle=90]{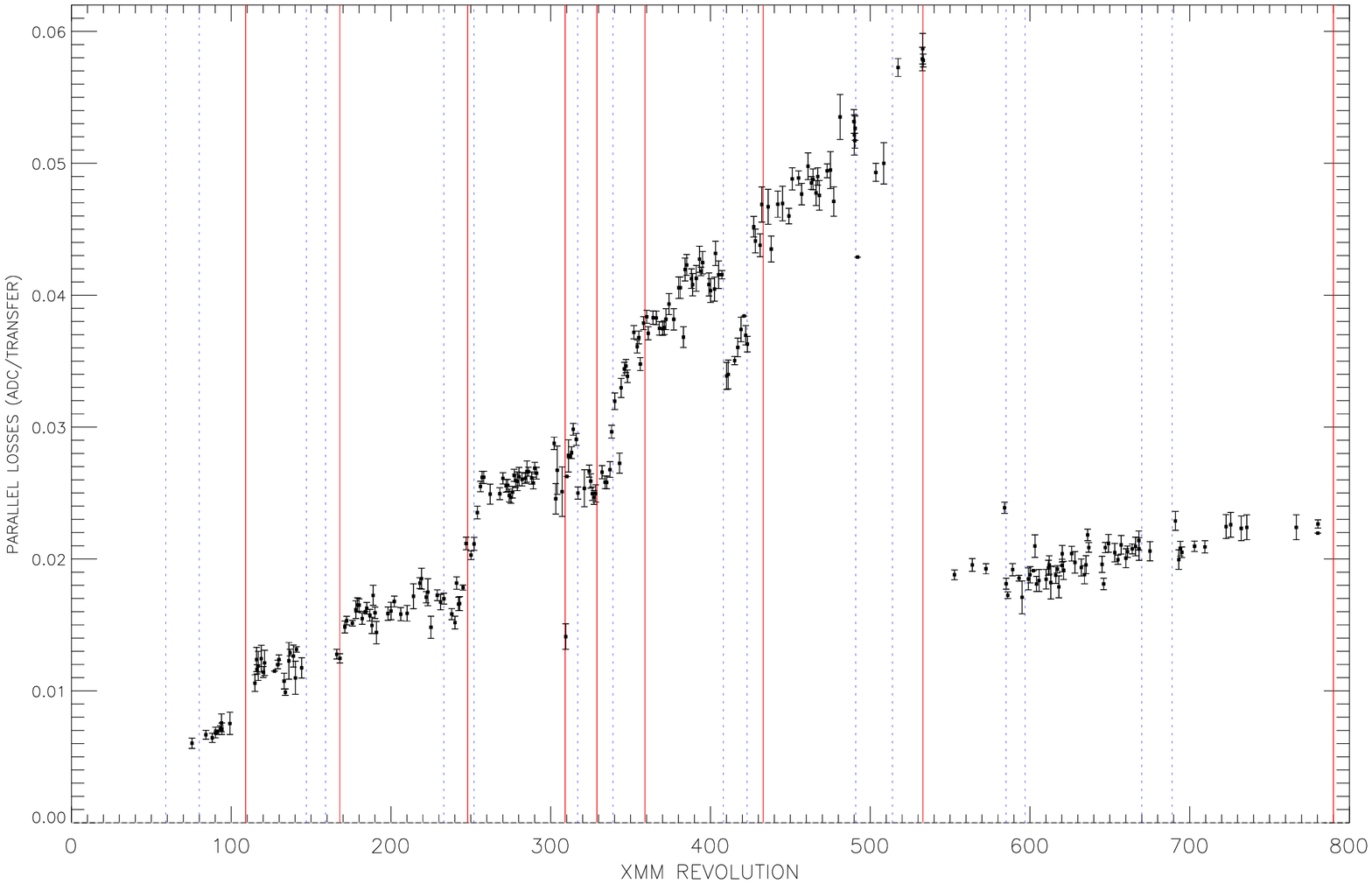}
   \end{tabular}
   \end{center} 
   \caption[MOS1 CTI] 
   { \label{fig:cti-mos} 
Parallel charge losses in units of ADC per transfer for the central CCD of MOS1 (left panel) and MOS2 (right panel) at Mn $\rm K_{\alpha}$. The gain of the CCDs are such that 1 ADC is approximately 3.25 eV. Vertical blue/dashed lines indicate spacecraft eclipse periods. Vertical red/solid lines indicate large solar flares at which discontinuities in the smoothly increasing trend in the CTI can be seen.}
   \end{figure}

CTI is monitored by using the $\rm Fe_{55}$ calibration source within the 
filter wheel assembly which produces strong lines at Al and 
Mn $\rm K_{\alpha}$. Fig.~\ref{fig:cti-mos} shows the evolution of the CTI 
of the central CCDs of MOS1 and MOS2 over 800 orbits since launch.

The vertical straight lines on the plots indicate spacecraft eclipse periods
and also particularly large solar flares at which discontinuites in the 
general increasing trend of the CTI can be detected. The upwards trend is
consistent with pre-launch calculations of the accumulated damage due to
passage of the radiation belts near perigee. The solar flares generate 
additonal huge fluxes of high energy protons which penetrate the shielding.

\begin{figure}[h]
   \begin{center}
   \begin{tabular}{c}
   \includegraphics[width=7cm,height=12cm,angle=90]{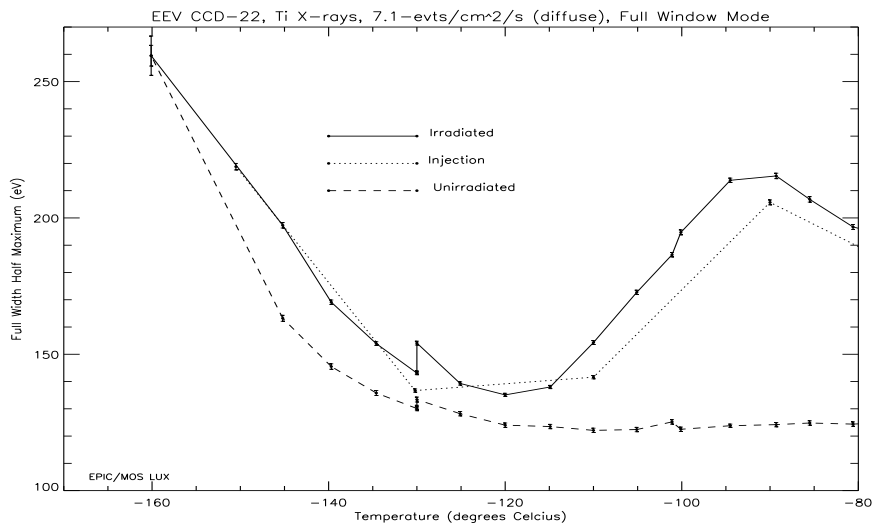}
   \end{tabular}
   \end{center}
   \caption[CTI Lab] 
   { \label{fig:ctilab} 
FWHM of the Ti $\rm K_{\alpha}$ line as a function of temperature. The lowest 
curve is before irradiation. The top two curves are after 5 years equivalent 
proton irradiation with (dotted line) and without (solid line) charge 
injection}
   \end{figure}

The large discontinuity and partial recovery of the CTI after Orbit 534 is 
due to cooling of the MOS cameras from a post-launch temperature of 
$-100^{\circ}$C down to $-120^{\circ}$C. The dependance of CTI on temperature 
was previously demonstrated by laboratory tests carried out on an EPIC-MOS 
type CCD which was irradiated with $\rm 2.5 \times 10^{8} \ cm^{-2}$ 10~MeV 
equivalent protons (Fig.~\ref{fig:ctilab}). This dosage was calculated to be 
typical of 5 years operation of XMM-Newton in space. Temperature affects the 
CTI because charge traps can be filled by electrons which allow other 
electrons to pass the trap safely. The electrons within the trap are released 
by thermal motion whose timescale is dependant on the temperature of the 
CCD\cite{Holland93}.

\begin{figure}[h]
   \begin{center}
   \begin{tabular}{c}
   \includegraphics[width=6cm,height=10cm,angle=270]{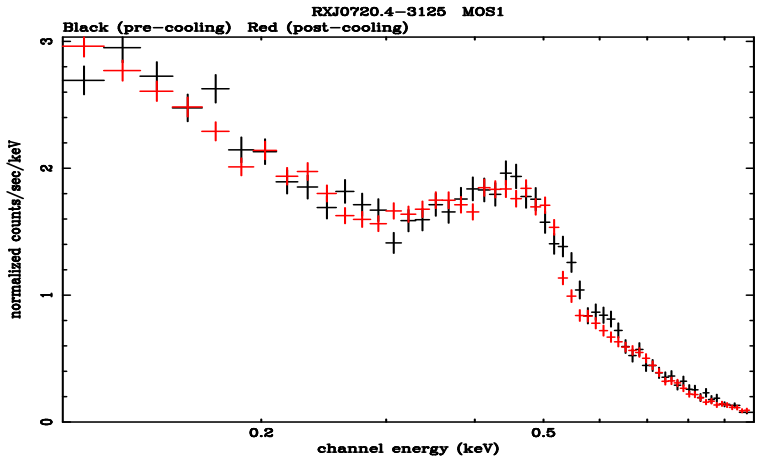}
   \end{tabular}
   \end{center}
   \caption[example] 
%>>>> use \label inside caption to get Fig. number with \ref{}
   { \label{fig:rxj0720} 
MOS1 spectra of the isolated NS RX~J0720.4-3215 before and after cooling of the MOS CCDs. The spectra have been corrected for the pre- and post-cooling CTI losses. No significant differences are observed between the spectra.}
   \end{figure}

After some in-orbit cooling tests, described in Ferrando et al. (2003)\cite{Ferrando03}, 
both cameras were cooled together during orbit 533. Observations of the isolated 
Neutron star RX~J0720.4-3125 and the Vela Supernova Remnant taken one orbit 
before and after cooling demonstrated no effects on the spectral response of 
the CCDs apart from those arising from the improvement in CTI 
(Fig.~\ref{fig:rxj0720}).

\subsection{Future CTI Correction}

CTI shifts the observed energies of X-rays to lower energies, but can be 
corrected for in software in a fairly straightforward manner. The algorithm 
to do this has been developed collaboratively by the instrument PI team and
the ESA Spacecraft Operations Centre (SOC) and is implemented within the 
Science Analysis System (SAS) used by observers to analyse XMM data. By 
design the absolute pre-launch CTI of the MOS 
CCDs was relatively modest\cite{Holland90} and the current CTI 
correction algorthim is based simply on the absolute number of transfers an 
event undergoes irrespective of which column(s) of the CCD the charge is 
deposited into.

Provision for calibrating individual columns of the CCDs was not part of the 
initial calibration effort. The relatively low CTI of the MOS CCDs meant that 
the additional computational overhead of calculating and correcting CTI for 
each of the 600 columns in 14 CCDs was considered prohibitively expensive. 
Five years is a long time in computing terms however, and the EPIC-MOS 
instrument team is now reinvestigating a column by column CTI correction 
algorithm. This should be especially beneficial to the study of 
line-dominated extended objects such as supernova remnants by improving
the residual spatial variation in resolution and line centroid determination.

   \begin{figure}[h]
   \begin{center}
   \begin{tabular}{c}
   \includegraphics[width=6cm,height=12cm,angle=90]{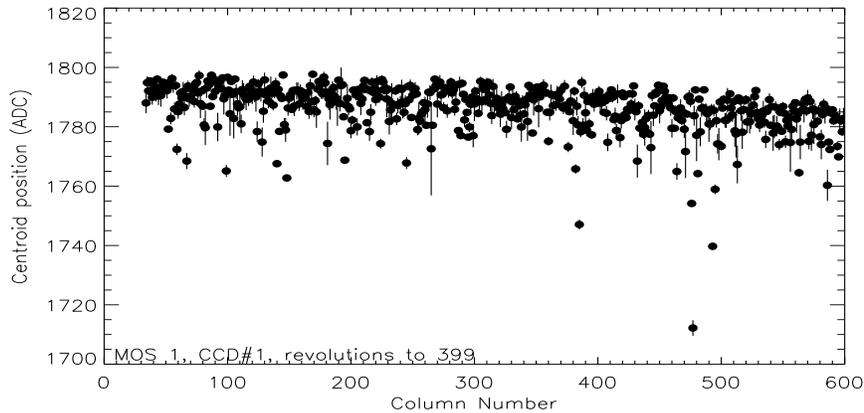}
   \end{tabular}
   \end{center}
   \caption[MOS1 Column] 
   { \label{fig:col-mos1} 
Mean centroid in ADC units of the uncorrected Mn~$\rm K_{\alpha}$ 
calibration line averaged over each column of the central CCD of MOS1. 
The data have been accumulated from closed calibration datasets summed 
between orbits 100 and 399. The slope in the average centroid from left to
right is due to CTI in the serial register.}
   \end{figure} 

An example of the column to column variation in the line centroid at 
Mn~$\rm K_{\alpha}$ is shown in Fig.~\ref{fig:col-mos1}. The most
extreme columns deviating from the average are currently masked as
bad within the SAS. 

\subsection{Resolution}
\label{sect:resolution}

\begin{figure}[h]
   \begin{center}
   \begin{tabular}{c}
   \includegraphics[width=6cm,height=8.3cm,angle=90]{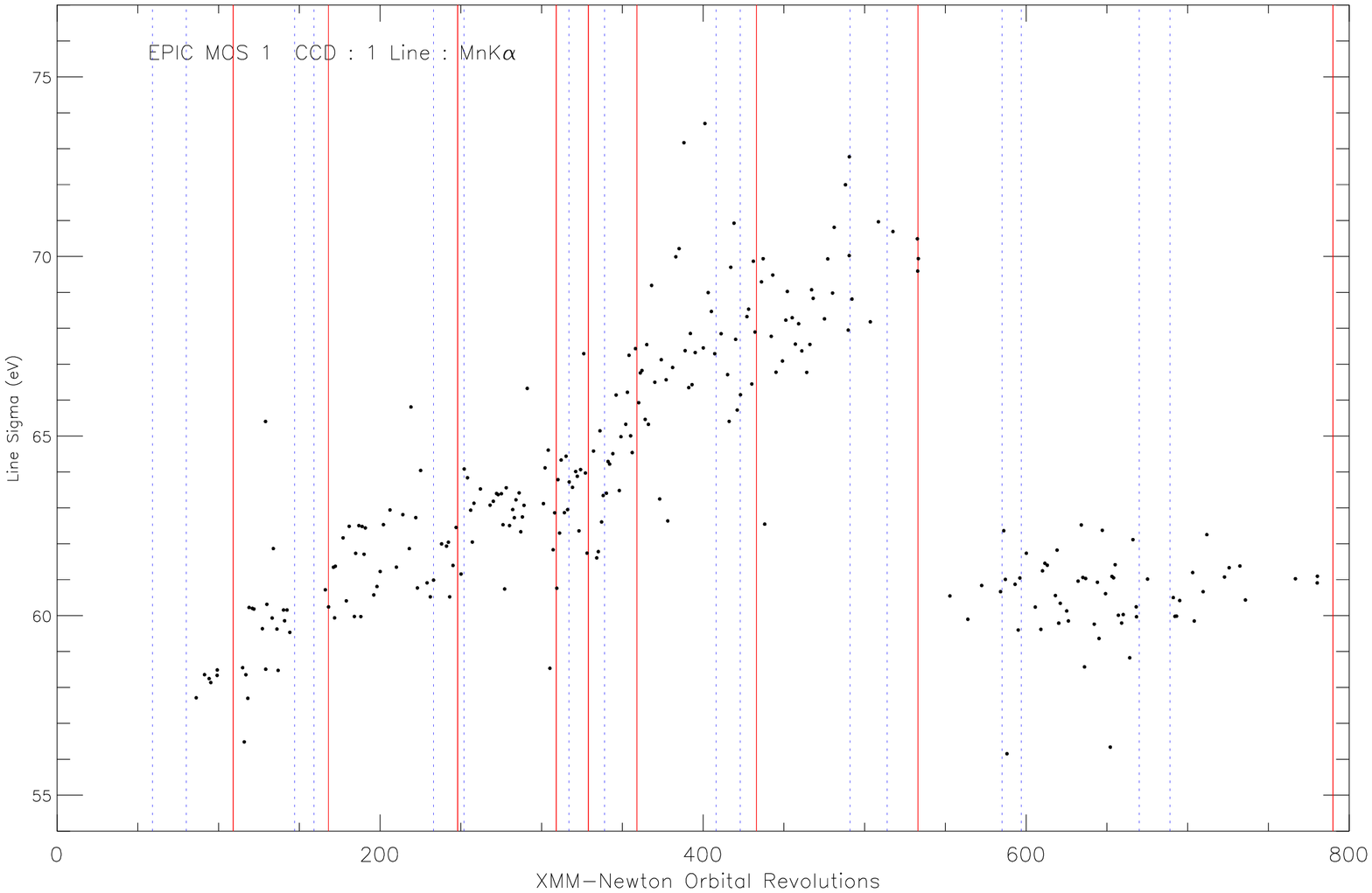}
   \includegraphics[width=6cm,height=8.3cm,angle=90]{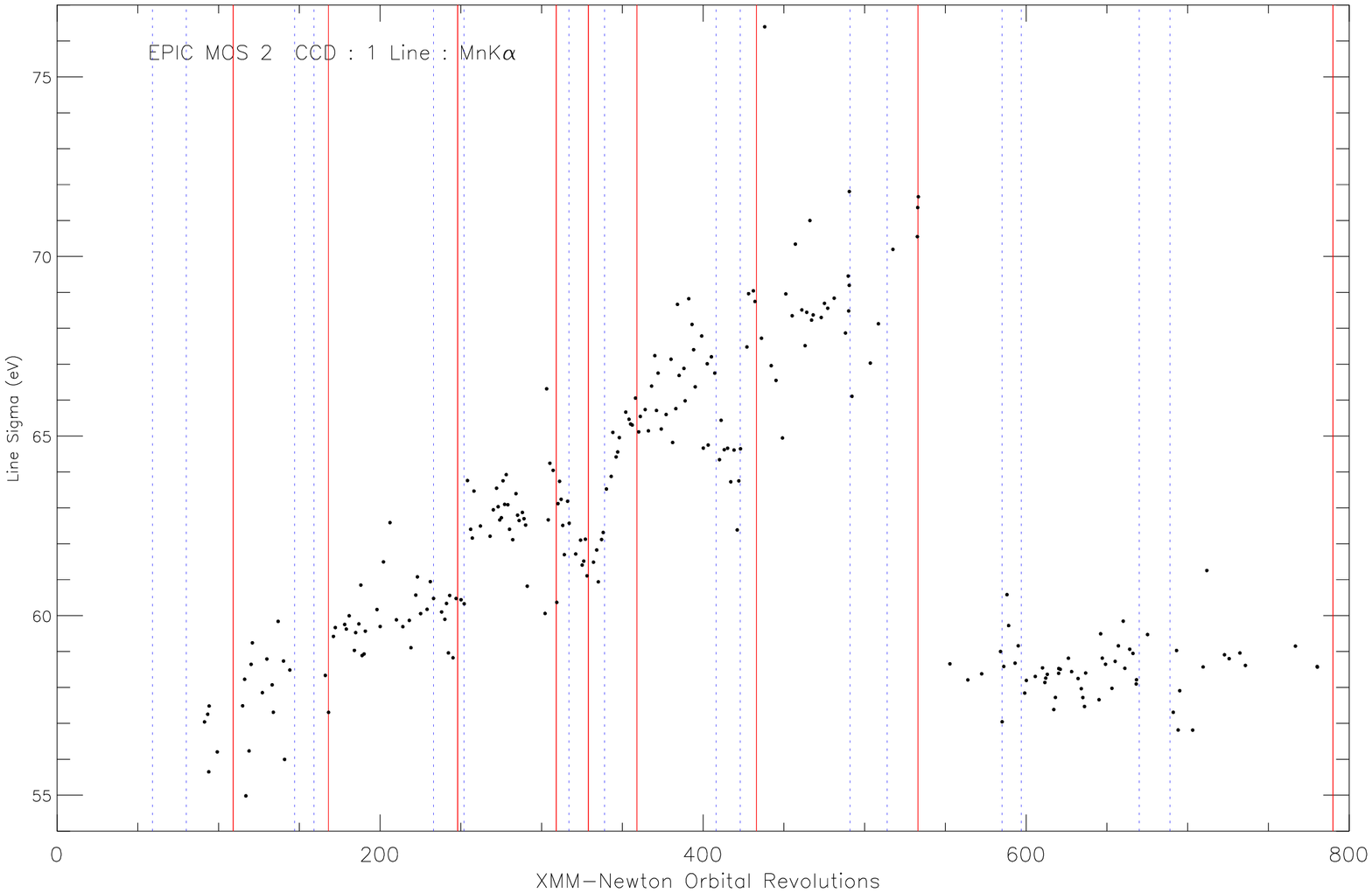}
   \end{tabular}
   \end{center}
   \caption[MOS1 Sigma] 
   { \label{fig:sig-mos} 
Evolution of the width (sigma) of the CTI-corrected Mn $\rm K_{\alpha}$ line for the central CCDs of MOS1 (left panel) and MOS2 (right panel)}
   \end{figure}

CTI losses are a statistical process which create an unrecoverable broadening 
of the energy resolution even if the best possible CTI correction algorithm
were to be applied. Fig.~\ref{fig:sig-mos} shows the evolution of the 
width of the CTI-corrected Mn $\rm K_{\alpha}$ 
calibration lines on the central CCDs of MOS1 and MOS2. During eclipse periods
the camera electronics are cooler and there are consequently small drops in the
line width.

Since cooling the rate of degradation in resolution and 
CTI (Sec.~\ref{charge-transfer-inefficiency}) has lessened signifcantly. 
This indicates that the CCDs appear to be more robust to radiation 
damage at the current operating temperature of $-120^{\circ}$C.

\subsection{The Soft Proton Problem}

The shielding designed to block high energy protons is avoided by soft 
protons (energies below around 300~keV) which are able to reach the CCDs 
because they are focussed by the grazing incidence mirrors. It was only 
after the launch of Chandra, a few months before XMM-Newton, that the 
problem of soft protons was recognised\cite{Prigozhin00}. These protons are 
stopped in less than a micron of material and are potentially very damaging 
for front-illuminated devices such as the EPIC-MOS CCDs as shown by
laboratory measurements\cite{Abbey01}. In comparison the 
EPIC-pn is self-shielded against soft proton damage as it is 
a 280~$\mu$m thick back-illuminated device.

The soft proton flux on the detectors is highly variable, varying by up to
four orders of magnitude\cite{Altieri03}. Flares occur most frequently
during, but are not restricted to, the approach to perigee at the end of 
orbits. The soft protons produce background events which are nominally 
indistinguishable from X-rays and are not well correlated with the 
measured flux in the radiation monitor on XMM-Newton which is only
sensitive to protons above 3.5 MeV. The original plan for protection of the 
cameras was for measurements taken with the radiation monitor to be used
to trigger the closing of the filter wheel during periods of
high radiation. This places 1.05~mm of aluminium shielding into the field 
of view. The insensitivity of the radiation
monitor to soft protons, however, meant that a new operational procedure
was developed to use a more reliable indicator of soft proton flaring
to trigger the filter wheel. Good indicators such as the discarded 
line counter in the pn camera\cite{Kendziorra00} and the X-ray count rate 
in peripheral CCDs are instrument mode dependant and are not always 
available throughout each orbit. It was found\cite{Kendziorra00} that 
these indicators are well correlated with an RGS\cite{denHerder01} 
parameter that records the number of events above threshold and is
continuously available.

Altieri (2003)\cite{Altieri03} has calculated the soft proton dosage
on the cameras after 3.25 years of operations to be about 
$\rm 4-6 \times 10^{5} \ protons \ cm^{-2}$ per EPIC-MOS camera and 
$\rm 1.2 \times 10^{6} \ protons \ cm^{-2}$ for EPIC-pn. The difference
arises because around half the proton flux is blocked by the reflection
gratings within the MOS telescopes and the more stringent safety 
procedures for MOS which often close the cameras before pn.

The observed rate of CTI damage appears to be close to pre-launch predictions
based on calculations of the high energy proton flux 
(Section~\ref{charge-transfer-inefficiency}). It would appear, therefore, that
that new operational safety procedures have been quite effective in 
minimising the effects of soft protons in this particular regard.

\subsection{Redistribution Function}

\begin{figure}[h]
   \begin{center}
   \begin{tabular}{c}
   \includegraphics[width=9cm,height=8cm]{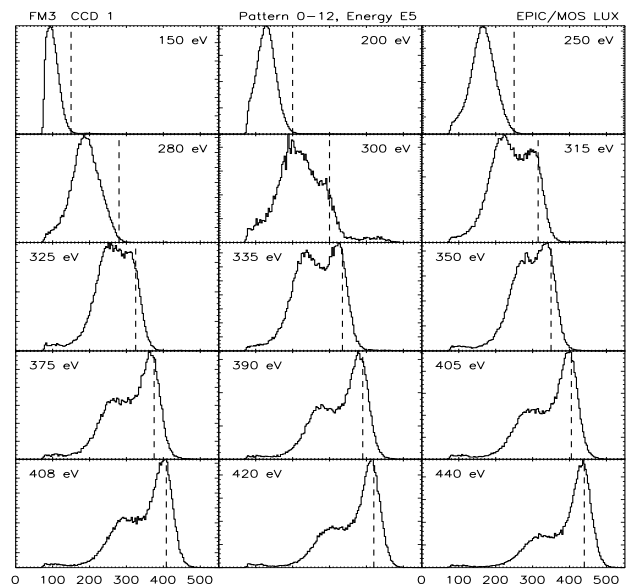}
   \end{tabular}
   \end{center}
   \caption[example] 
%>>>> use \label inside caption to get Fig. number with \ref{}
   { \label{fig:m1-orsay} 
Ground calibration data from the central CCD of MOS1 (Flight Model 3) showing 
the observed spectra (after gain and CTI correction) for a series of
monochromatic input energies. The evolution of the surface 
loss {\it shoulder} is evident.}
   \end{figure}

The redistribution function (rmf) for a monochromatic input has a rather 
complex shape which increasingly deviates from an {\it ideal} Gaussian 
distribution towards lower energies. Fig.~\ref{fig:m1-orsay} shows ground 
calibration data taken at the Orsay synchrotron\cite{Dhez97} of the 
central CCD of MOS1 (Flight Model Number 3) at a range of monochromatic 
input energies from 150 
up to 440 eV. As the input energy decreases a {\it shoulder} below 
the main photopeak increases in relative strength until it completely 
dominates the rmf. 

\begin{figure}[h]
   \begin{center}
   \begin{tabular}{c}
   \includegraphics[width=7cm,height=12cm,angle=90]{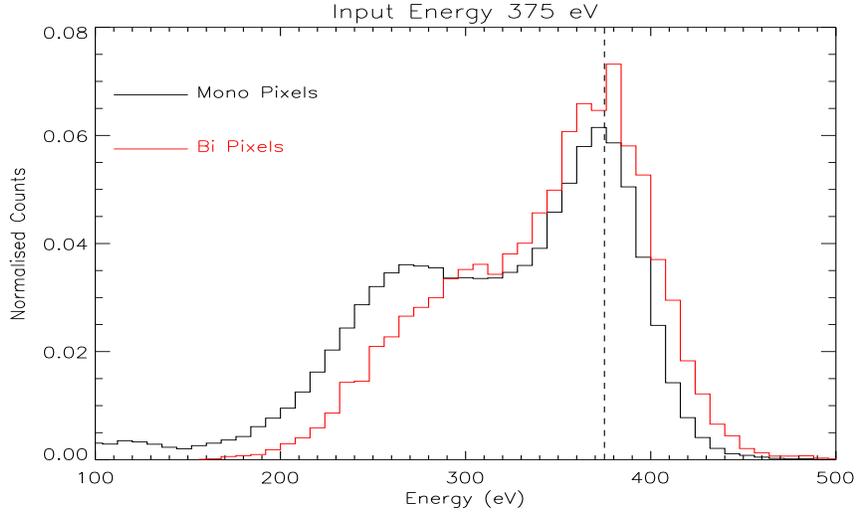}
   \end{tabular}
   \end{center}
   \caption[example] 
%>>>> use \label inside caption to get Fig. number with \ref{}
   { \label{fig:mono-bi} 
Ground calibration data from the central CCD of MOS1 showing normalised mono-pixel (black/bold) and bi-pixel spectra for a monochromatic beam of 375 eV. The bi-pixel spectrum has a smaller surface loss shoulder than the mono-pixel spectrum.}
   \end{figure}

The origin of the {\it shoulder} is thought to be incomplete charge 
collection for X-rays absorbed near the surface layer caused by
an inversion in the surface potential\cite{Short98}. The effect is much 
stronger for mono-pixels than bi-pixels (Fig.~\ref{fig:mono-bi}). The MOS
CCDs have an etched open phase which was designed to increase the low energy
quantum efficiency and the inversion in the surface potential is likely to be 
stronger in this open phase: At low energies the majority of mono-pixel events 
are absorbed within the open phase whereas bi-pixel events arise from photons 
absorbed near pixel boundaries which are partially covered by
the surface electrode structure\cite{Hiraga01}. 

\begin{figure}[h]
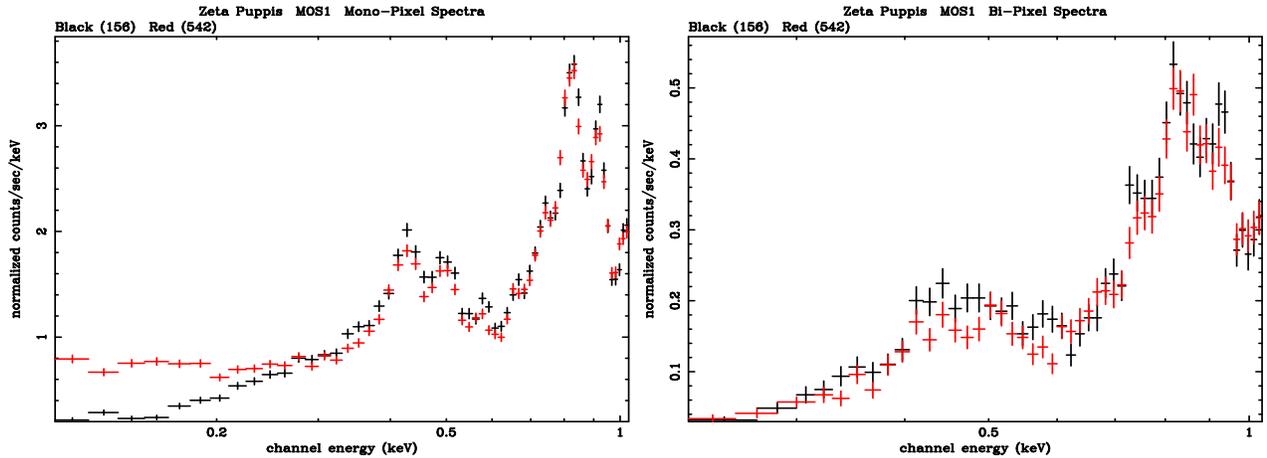

   \begin{center}
   \begin{tabular}{c}
   \includegraphics[width=6cm,height=8.3cm,angle=270]{zeta-m1-mono.ps}
   \includegraphics[width=6cm,height=8.3cm,angle=270]{zeta-m1-bipx.ps}
   \end{tabular}
   \end{center}
   \caption[example] 
%>>>> use \label inside caption to get Fig. number with \ref{}
   { \label{fig:m1-zeta} 
Mono-pixel (left panel) and bi-pixel (right panel) background 
subtracted spectra of the bright O star $\zeta$~Puppis taken in orbits 156 (black/bold) and 542.}
   \end{figure}

\begin{figure}[h]
   \begin{center}
   \begin{tabular}{c}
   \includegraphics[width=6cm,height=8.3cm,angle=270]{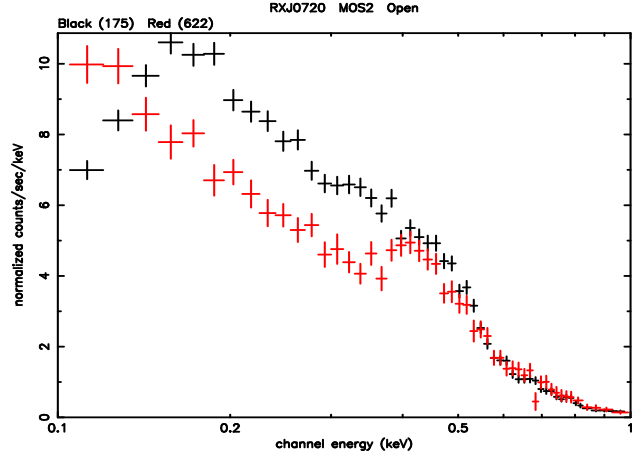}
   \end{tabular}
   \end{center}
   \caption[example] 
%>>>> use \label inside caption to get Fig. number with \ref{}
   { \label{fig:m2-open} 
Mono-pixel background subtracted MOS2 spectra of the isolated 
neutron star, RXJ0720.4-3125, taken during orbits 175 (black/bold) and 622}
   \end{figure}

Analysis of astronomical targets has shown that the shape of the rmf at low 
energies is not constant but has evolved throughout the mission. Both cameras
have been affected as illustrated by Fig.~\ref{fig:m1-zeta} which shows two 
sets of MOS1 spectra of the O star, $\zeta$~Puppis taken at different epochs, 
and Fig.\ref{fig:m2-open} which shows MOS2 spectra of the isolated Neutron
star, RX~J0720.4-3125.

$\zeta$~Puppis is an O star with strong emission lines at Nitrogen and 
Oxygen\cite{Kahn01}. The observed flux in the MOS seen at energies below the 
Nitrogen lines at around 430 eV 
is dominated by redistribution from the lines and residual continuum at 
higher energies. The observations of $\zeta$~Puppis show a difference in the 
mono-pixel spectra which is not reflected in the bi-pixel spectra. Similar 
results are seen for MOS2. These results indicate a change in the 
redistribution function which must be strongest in the open phase of the CCD.

The INS RXJ0720.4-3125 is now known to have a black-body like spectrum which
has slightly hardened over the lifetime of XMM\cite{deVries04}. The 
differences in the observed MOS2 spectra (also seen in MOS1) can be 
attributed to a combination of this spectral evolution and a change in the
rmf of the CCDs. The {\it increase} in flux below 200 eV for example must
be due to an rmf change as the flux in this band is dominated by events
redistributed downwards from higher energies.

\begin{figure}[h]
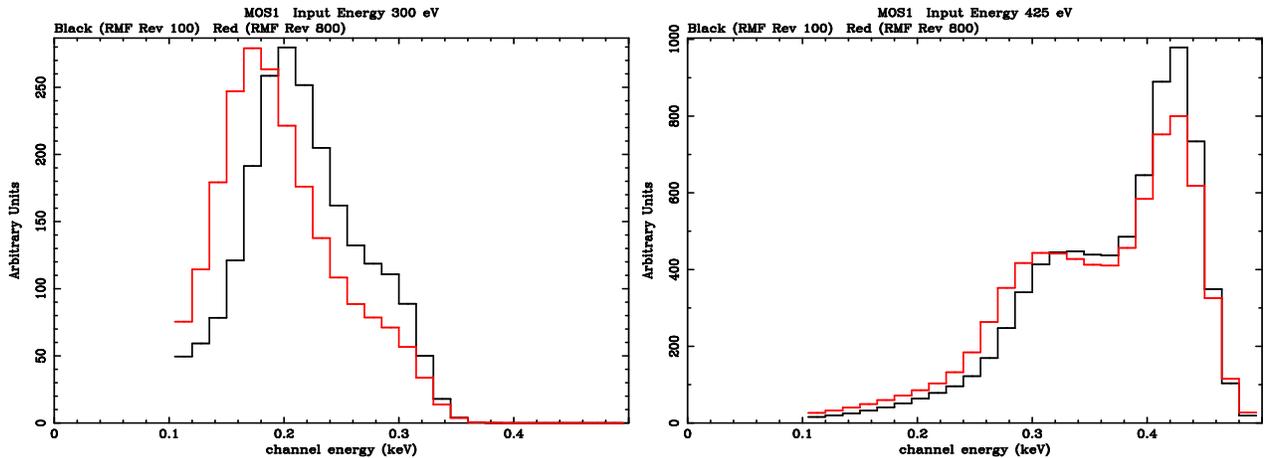

   \begin{center}
   \begin{tabular}{c}
   \includegraphics[width=6cm,height=8.3cm,angle=270]{mos1-300ev.ps}
   \includegraphics[width=6cm,height=8.3cm,angle=270]{mos1-425ev.ps}
   \end{tabular}
   \end{center}
   \caption[example] 
%>>>> use \label inside caption to get Fig. number with \ref{}
   { \label{fig:example-rmf} 
The shape of the MOS1 redistribution function for input energies of 300
and 425 eV. The rmfs have been generated for observations appropriate to
orbits 100 (black/bold) and 800.}
   \end{figure}

The data are consistent with the degree of charge loss within the surface loss
zone on the open phase increasing with time. The culprit is likely to be
the soft protons which have a higher probability of reaching the active region
in the open rather than the closed phase. The surface charge loss is 
modelled analytically by a simple linear function which describes the 
fractional charge loss as a function of depth with the active silicon. The
parameters of the function are tuned based on the in-orbit observations of 
line dominated sources such as $\zeta$~Puppis. Epoch dependant rmfs are now 
generated automatically by the SAS (Version 6.0 and above). 
Fig.~\ref{fig:example-rmf} shows the shape of the rmf for two input energies 
of 300 and 425 eV appropriate for typical observations in orbits 100 and 800.

%%%%%%%%%%%%%%%%%%%%%%%%%%%%%%%%%%%%%%%%%%%%%%%%%%%%%%%%%%%%%
\acknowledgments     %>>>> equivalent to \section*{ACKNOWLEDGMENTS}       
 
The authors would to like to thank the many people within ESA and the PI 
instrument teams who have contributed towards making XMM-Newton an outstanding
success. The support of PPARC, CEA and ESA in funding large parts of the 
post-launch calibration activities is gratefully acknowledged.

%%%%%%%%%%%%%%%%%%%%%%%%%%%%%%%%%%%%%%%%%%%%%%%%%%%%%%%%%%%%%
%%%%% References %%%%%

\bibliography{report}   %>>>> bibliography data in report.bib
\bibliographystyle{spiebib}   %>>>> makes bibtex use spiebib.bst

\end{document}